\documentclass{conm-p-l}
  [1995/01/05 v2.2e AMS font definitions%
  ]
\DeclareFontFamily{OT1}{wncyss}{}

\DeclareFontShape{OT1}{wncyss}{m}{n}{
   <5> <6> <7> <8> <9> <10> <10.95> <12> <14.4> <17.28> <20.74> <24.88> 
wncyss10  }{}

\DeclareFontFamily{OT1}{wncyr}{}
\DeclareFontShape{OT1}{wncyr}{m}{n}{
   <5> <6> <7> <8> <9> <10> <10.95> <12> <14.4> <17.28> <20.74> <24.88> wncyr10
  }{}
\DeclareFontShape{OT1}{wncyr}{bx}{n}{
   <5> <6> <7> <8> <9> <10> <10.95> <12> <14.4> <17.28> <20.74> <24.88> wncyb10
  }{}
\DeclareFontShape{OT1}{wncyr}{m}{it}{
   <5> <6> <7> <8> <9> <10> <10.95> <12> <14.4> <17.28> <20.74> <24.88> wncyi10
  }{}

\DeclareFontShape{OT1}{wncyr}{m}{sc}{
   <5> <6> <7> <8> <9> <10> <10.95> <12> <14.4> <17.28> <20.74> <24.88> 
wncysc10  }{}

\newtheorem{theorem}{Theorem}[section]

\theoremstyle{definition}
\newtheorem{definition}[theorem]{Definition}

\newtheorem{thm}{Theorem}[section]

\theoremstyle{definition}
\newtheorem{df}[theorem]{Definition}

\theoremstyle{remark}

\numberwithin{equation}{section}

\newcommand{\cal}{\mathcal}

\begin{document}
\title{The (secret?) homological algebra of the \bv approach 	}

\author{Jim Stasheff }
\address{Math Dept., University of North Carolina, Chapel Hill NC 27599-3250}
\email{jds@math.unc.edu }
\thanks{Research supported in part by NSF
grants DMS-8506637, DMS-9206929, DMS-9504871 and a W.R. Kenan Research
and Study Leave
from the University of North Carolina.  Thanks are also due to
the University of Pennsylvania for hospitality of many summers
and the current year.	}


\date{August 20, 1997}

\dedicatory{In memory of Chih-Han Sah 1934-1997}

\def\BV{\mathcal{BV}}
\def\lh{\hbox to 15pt{\vbox{\vskip 6pt\hrule width 6.5pt height 1pt}
\kern -4.0pt\vrule height 8pt width 1pt\hfil}} 
\def\blob{\mbox{$\;\Box$}}
\def\qed{\hbox{${\vcenter{\vbox{\hrule height 0.4pt\hbox{\vrule width
0.4pt height 6pt \kern5pt\vrule width 0.4pt}\hrule height 0.4pt}}}$}}
\def\varbic{variational bicomplex\ }
\def\CJ{Loc(E)}

\hyphenation{as-so-ci-a-he-dra back-ground com-mu-ta-tive
com-pac-ti-fi-ca-tion
com-plex cor-re-spond-ing de-for-ma-tion Geo-met-ri-cal-ly ho-mot-o-py
iden-ti-ties ori-en-ta-tion per-mu-ta-tions Pre-print pro-jec-tion re-spect
re-spec-tive-ly struc-tures sub-op-er-ad Hoch-schild}

\newcommand{\mathbold}{\mathbf}
\newcommand{\DM}{De\-ligne-Knud\-sen-Mum\-ford\ }
\newcommand{\hl}{ho\-mot\-o\-py Lie\ }
\newcommand{\coh}{co\-ho\-mol\-o\-gy}
\newcommand{\DKM}{De\-ligne-Knud\-sen-Mum\-ford\ }
\newcommand{\nc}{\mathbf{C}}
\newcommand{\nr}{\mathbold{R}}
\newcommand{\nl}{\mathbold{L}}
\newcommand{\nz}{\mathbold{Z}}
\newcommand{\bw}{\bigwedge}
\newcommand{\bv}{Ba\-ta\-lin-Vil\-ko\-vi\-sky\ }
\def\16N{d\lbrack\phi_1,\dots,\phi_N\rbrack+ \Sigma_{i=1}^N
      \pm \lbrack\phi_1,\dots, d\phi_i,\dots,\phi_N\rbrack=
      \hspace{1.25in}\\\hspace{2in}
      \Sigma^{N-1}_{Q=2}\pm \lbrack\lbrack\phi_{i_1},\dots,
      \phi_{i_Q}\rbrack,\phi_{i_{Q+1}},\dots,\phi_{i_N}\rbrack}
\def\L{\cal L}
\def\x{\times}
\def\Cn{F(S^1,n)}
\def\C2{F(S^1,2)}
\def\C3{\overbar F(S^1,3)}
\def\C4{\overbar F(S^1,4)}
\def\CnR{F(R^3,n)}
\def\CcR{$\overbar F(R^3,n)$}
\def\Cc{$\overbar F(S^1,n)$}
\def\Ccn{\overbar F(S^1,n)}
\def\overbar{\overline}
\def\ip{\coprod}
\def\overcirc{\stackrel{\circ}{\Delta}}
\def\overcircI{\stackrel{\circ}{I}}
\newcommand{\D}[1]{{\cal D}(#1)}
\newcommand{\F}[1]{{\cal F}(#1)}

\newcommand{\ad}{\operatorname{ad}}
\newcommand{\sym}{\operatorname{sym}}
\newcommand{\bdel}{\delta}
\newcommand{\aaa}{\alpha}
\newcommand{\bb}{\beta}
\newcommand{\ggg}{\gamma}
\newcommand{\dd}{\delta}
\newcommand{\rr}{r^a_\alpha}
\newcommand{\bfb}{\bold{b}}
\newcommand{\bfv}{\bold{v}}
\newcommand{\bft}{T}
\newcommand{\codim}{\operatorname{codim}}
\newcommand{\CP}[1]{\mathbold{C}\mathbold{P}^{#1}}
\newcommand{\edges}[1]{\operatorname{edges}(#1)}
\newcommand{\End}[1]{\cal{E}nd\,(#1)\,}
\newcommand{\gh}{\operatorname{gh}}
\newcommand{\HH}{\cal{H}} 
\newcommand{\Hom}{\operatorname{Hom}}
\newcommand{\Hr}{\HH_{\operatorname{rel}}}
\newcommand{\Hrr}{H_{\operatorname{rel}}}
\newcommand{\hr}{\HH^{\operatorname{rel}}}
\newcommand{\id}{{\operatorname{id}}}
\newcommand{\Int}{\operatorname{int}}
\newcommand{\Mg}[1]{\cal{M}_{g, #1}}
\newcommand{\Mz}[1]{\cal{M}_{0, #1}}
\newcommand{\My}[1]{\cal{M}_{0, #1 +1}}
\newcommand{\m}{_{\operatorname{m}}}
\newcommand{\map}[2]{\operatorname{Map} (#1,#2)}
\newcommand{\Mc}[1]{\overline{\cal{M}}_{#1}}
\newcommand{\Mgc}[1]{\overline{\cal{M}}_{g, #1}}
\newcommand{\Mn}{\cal{M}_{n}}
\newcommand{\Mgn}{\cal{M}_{g,n}}
\newcommand{\Mzn}{\cal{M}_{0,n+1}}
\newcommand{\Mgnc}{\overline{\cal{M}}_{g,n}}
\newcommand{\N}[1]{\cal{N}_{#1}}
\newcommand{\Nc}[1]{\underline{\cal{N}}_{#1}}
\newcommand{\OR}{\operatorname{or}}
\newcommand{\out}{\operatorname{out}}
\newcommand{\PGL}{\operatorname{PGL}(2,\nc)}
\newcommand{\p}[1]{\cal{P}_{#1}}
\newcommand{\Pt}{\operatorname{pt}}
\newcommand{\V}[1]{\cal{V}_{#1}}
\newcommand{\Vir}{\operatorname{Vir}}
\newcommand{\X}[1]{\underline{\cal{M}}_{#1}}
\newcommand{\Xg}[1]{\underline{\cal{M}}_{g,#1}}
\newcommand{\CHA}{$C_{\infty}$-algebra}
\newcommand{\SHA}{$A_{\infty}$-algebra}
\newcommand{\ci}{\cite}
\newcommand{\DD}{\Delta}
\newcommand{ \BBvD}{Burgers, Behrends and van Dam}
\newcommand{\doubleoverbar}{$\tilde L$}
\newcommand{\smooth}{$C^\infty(M)$}

\def\L{L_\infty}
\def\LL{$$\overbar L = L + \phi^*_ir^i_\alpha C^\alpha
+ (C_{\ggg}^*f^{\ggg}_{\alpha\bb} + \phi^*_i\phi^*_j\4mu4)
C^\alpha C^\bb$$}
\def\SS{$\overbar S =\int_M\overbar L dvol_M$}
\def\s{s=(\overbar L,\ \ ) }
\def\umu{u^{i\overbar \mu}}
\def\4mu4{\nu^{a}_{\alpha\beta}}

\begin{abstract}After a brief history of `cohomological physics',
the \bv complex is given a revisionist presentation as homological
algebra, in part classical, in part novel.  Interpretation of the
higher order terms in the extended Lagrangian is given as higher
homotopy Lie algebra and via deformation theory.  Examples are given 
for higher spin particles and closed string field theory.
\end{abstract}

\maketitle
\def\CJ{Loc(E)}
\def\be{\begin{eqnarray}}
\def\ee{\end{eqnarray}}

\font\cyr=wncyr10

\newcommand{\ui}{{\rm \v{\cyr i}}}

\hyphenation{gla-v-no-go gra-du-i-ro-va-n-no gra-du-i-ro-va-n-noe
ig-ra-et
ime-et kva-d-ra-tov li-e-vy ma-te-ma-tiche-s-kih  mno-go-ob-ra-zii
od-no-ro-d-ny ope-rad
pe-re-s-ta-no-v-ke slu-zhit stru-k-tur stru-k-tu-ru stru-k-tu-ro
to-zh-destv to-zh-de-st-vo voz-ro-zh-de-niyu Bat-a-lina Vil-ko-vy-ssko-go}

\newcommand{\h}{Hoh\-shilp1\-da}
\newcommand{\hc}{kom\-pleks Hoh\-shilp1\-da}
\newcommand{\hg}{go\-mo\-to\-pi\-che\-s\-kaya algebra Gerstenhabera\ }

\newcommand{\ca}{\cal{C}^\bullet (A)}
\newcommand{\CA}[1]{\cal{C}^{#1} (A)}
\newcommand{\hha}{\operatorname{HH}^\bullet (A)}

\renewcommand{\thesubsection}{\arabic{subsection}}


\begin{cyr}

Prezhde vsego, ya hotel by poblagodaritp1 organizatorov e1to\ui \
konferentsii, v pervuyu ocheredp1, Mikhaila Mikha{\ui}lovicha i Iosifa
Semenovicha.  Ya rad vstreti\-tp1\-sya s russkimi i ukrainskimi
matematikami i fizikami.  Ya nadeyusp1, chto v budushchem mnogie iz
vas smogut posetitp1 Chepl Hil.

Segodnya ya hotel by obsuditp1 svyazi mezhdu nashimi rabotami na
Zapade i vashimi rabotami v Rossii, osobenno rabotami Batalina i
Vilkovyskogo.  Ya dumayu, chto zdesp1 ya, vozmozhno, budu
rasskazyvatp1 vam kak ispolp1zovatp1 samovar.

Kazhdy{\ui}\ lyubit yazyk svoe{\ui}\ strany.  Poe1tomu ya budu govoritp1
po-angli{\ui}ski.

\end{cyr}
The following exposition is based in large part on work
with   Glenn Barnich (ITP, Berlin and Universit\'e Libre de
Bruxelles) and Tom Lada and Ron Fulp of NCSU (The Non-Commutative
State University). It is closely related to several other talks
at this conference.

I am particularly happy to see `Cohomological Physics' in the title of this 
conference.  I first referred to cohomological physics in the context of 
anomalies in gauge theory, cf. my work with Bonora, Cotta-Ramusino and
Rinaldi \cite{BCRSI,BCRSII}, more than a decade ago.
(Some people thought that phrase was a bit much.)  The cohomology referred to 
there was that of differential forms (the de Rham complex). 
Differential forms were implicit in physics at least as far back as 
Gauss (1833)
(cf. his electro-magnetic definition of the linking number \ci{gauss}), and more
visibly in Dirac's magnetic monopole (1931) \cite{Dirac:mm}, which lived in a 
$U(1)$ bundle over $\mathbold{R}^3-0$.  The magnetic charge was given by the 
first Chern number; for magnetic charge $1$, the monopole lived in the 
Hopf bundle, introduced that same year by Hopf \cite{hopf}, though it seems 
to have taken some decades for that coincidence to be recognized 
\cite{greub-petry}.
Thus were characteristic classes (and by implication the cohomology of
Lie algebras and of Lie groups) secretly introduced into physics.

Cohomological physics had a major breakthrough with 
the `ghosts' introduced by  Fade'ev and Popov \cite{fad-pop}. 
These were incorporated into  what came to be known as BRST cohomology 
(Becchi-Rouet-Stora \cite{BRS} and, here in Russia, Tytutin \cite{tyutin})
and which was applied to a variety
of problems in mathematical physics.  There the  ghosts were reinterpreted by
Stora \cite{stora} and others in terms of the Maurer-Cartan forms in the
case of a finite dimensional Lie group and more generally as generators
of the Chevalley-Eilenberg complex \cite{CE} for Lie algebra cohomology.

Group theoretic cohomology had already appeared in the work of Bargmann 
\cite{barg:lor,barg:cont} on extensions of the Galilean, Lorentz   and    
de Sitter groups. 

 Although I was unaware of it at the time, deformation theory (cf. Gerstenhaber
\cite{gerst:coh}) 
also provided a bridge between the kind of homotopy theory I was doing and 
mathematical physics via deformation quantization (Bayen, Flato, Fronsdal,
Lichnerowicz, Sternheimer \cite{bfflsI,bfflsII}), though the cohomological aspects 
were of minor importance in that application originally.

What I did become aware of next, thanks to Henneaux \cite{Henn:physrep}
and Browning and McMullen \cite{BM}, 
was the cohomological reduction of constrained Poisson algebras, specifically 
the Batalin-Fradkin-Vilkovisky approach \cite{BF,FF,FV}, which extended BRST by 
reinventing the 
Koszul-Tate resolution of the ideal of constraints and producing a synergistic 
combination of both kinds of cohomology.  Here it was that I saw the esential 
features of a strong homotopy Lie algebra ($L_\infty$-algebra).

On the other hand, cohomology was the essence of the \varbic approach to 
variational problems (Vinogradov \cite{vino},Tsujishita \cite{tsujishita}, 
Anderson \cite{ian:contemp,ian:book}) in which Lagrangians and 
the Euler-Lagrange equations were central. Here we already encounter 
cohomological physics in the guise of differential forms on the jet bundle.
As I learned here at this conference, Krasil'shchik \ci{krasil} has
developed the relation of deformation
theory to the \varbic approach.  An essential feature is the rich algebraic 
structure of various generalizations of Poisson brackets, including Schouten,
Schouten-Nijehnhuis, Gerstenhaber and  Nijenhuis-Richardson brackets.

The Batalin-Vilkovisky machinery for Lagrangian field theory is often
presented in terms of fields which are functions on some manifold or
sections of some bundle (cf. Henneaux's talks at this conference).  
The alternative approach which I will adopt 
for this conference reworks the machinery using the jet bundle approach. 
For years I've been tantalized by 
the idea of combining the BV machinery with the \varbic and had originally 
planned to talk on that here, but instead, with  my coauthors (Barnich, Fulp 
and Lada), I have been concentrating on elaborating the BV approach to 
Lagrangian field theory, especially the higher homotopy aspects.
We work with this in terms of the Euler-Lagrange 
complex which occurs at the edges of the \varbic. 
I will make some remarks about the combination of the anti-field, 
anti-bracket machinery with the \varbic; such a combination will be directly 
relevant  in the context of constraints and/or symmetries.
Such a combination has already appeared in works of McCloud \cite{mccloud}
and Barnich-Henenaux \cite{bh:isoms}.

Hopefully this will inspire more of you to follow this route in
delving further into the \varbic.

Thus we see a rather intricate interweaving of several
kinds of cohomology being brought to bear on problems in physics.
I will not try to describe the whole web, but rather will follow one
strand with just a few comments as others intersect it.
The Batalin-Vilkovisky approach
\cite{bv:antired,bv:closure,BV3} to quantizing particle Lagrangians and
Lagrangians of string field theory involves the rubric of anti-fields
as well as ghosts and an `anti-bracket', first introduced essentially
by Zinn-Justin \cite{zj,zj:qftbook} in another notation.
A revisionist view of the Batalin-Vilkovisky machinery recognizes 
parts of it  as a reconstruction of homological algebra \cite {ht} 
with some powerful new ideas undreamt of in that discipline. 
The `standard construction' is the \bv  complex,
and again we see the signature of $L_\infty$-algebras.

The `quantum' Batalin-Vilkovisky master equation has the form of the 
Maurer-Cartan equation for a flat connection, while the `classical' 
version has the form of the integrability equation of deformation theory. 
Just as the Maurer-Cartan equation makes sense in the context of Lie
algebra cohomology, so the \bv master equation has an interpretation in
terms of $L_\infty$-algebras.
Particularly interesting examples are provided by Zwiebach's closed
string field theory \cite{z:csft} and higher spin particles \cite{burgers:diss,
BBvD:three}.
\tableofcontents
\contentsline {section}{\tocsection {}{1}{The jet bundle setting for Lagrangian field theory}}{3}
\contentsline {section}{\tocsection {}{2}{Anti-fields as Koszul generators}}{5}
\contentsline {section}{\tocsection {}{3}{Noether identities and Tate generators}}{5}
\contentsline {section}{\tocsection {}{4}{Ghosts and the anti-bracket}}{7}
\contentsline {section}{\tocsection {}{5}{The Batalin-Vilkovisky complex $(\mathcal {BV},s_\infty )$}}{7}
\contentsline {section}{\tocsection {}{6}{The BV version of the variational bicomplex\ }}{9}
\contentsline {section}{\tocsection {}{7}{The Master Equation and Higher Homotopy Algebra}}{9}
\contentsline {section}{\tocsection {}{8}{Deformation Theory and the Master Equation in Field Theory}}{11}
\contentsline {section}{\tocsection {}{9}{Quantization and other puzzles}}{12}
\contentsline {section}{\tocsection {}{}{References}}{13}

\section{The jet bundle setting for Lagrangian field theory}

Let us begin with a space  $\Phi$  of {\bf fields} regarded as the space
of sections of some bundle $\pi:E \to M$.  For expository and coordinate
computational purposes, I will assume $E$ is a trivial vector bundle and
will write a typical field as $\phi = (\phi^1,\dots,\phi^k):M\to
\nr^k$. In terms of local coordinates, 
the base manifold M is locally $\nr^n$ with
coordinates $x^i, i=1,\dots,n$ and the fibre is $\nr^k$ with coordinates $u^a,
a=1,\dots,k$.  We `prolong' this bundle to create the associated jet
bundle $J=J^\infty E  \to E \to M$ which is an infinite dimensional
vector bundle with coordinates $u^a_I$ where $I=i_1\dots i_r$ is a
symmetric multi-index (including, for $r=0$,  the empty set of indices,
meaning just $u^a$).  The notation is chosen to bring to mind the mixed
partial derivatives of order $r$. Indeed, a section of $J$ is the
(infinite) jet $j^\infty \phi$ of a section $\phi$ of $E$ if, for all
$r$, we have $u^a_I\circ j^\infty \phi=
\partial_{i_1}\partial_{i_2}...\partial_{i_r} \phi^a$
 where $\phi^a = u^a\circ \phi$ and $\partial_i =
\partial/\partial x^i$.

\begin{definition}\rm
A {\bf local function} $L(x,u^{(p)})$ is  the pullback of a smooth function on some
finite jet bundle $J^pE$, i.e.  a composite $J \to J^pE \to \nr$.
In local coordinates, it is a smooth function of the $x^i$ and the $u^a_I$, 
where the order $|I| = r$
of the multi-index $I$ is less than or equal to some integer $p$.
The {\bf space of local functions} will be denoted $\CJ$.
\end{definition}

\begin{definition}\rm A {\bf local functional}
\begin{eqnarray}
{\mathcal L}[\phi]=\int_M  L(x,\phi^{(p)}(x)) d{vol}_M
= \int_M (j^\infty \phi)^*  L(x,u^{(p)}) d{vol}_M
\end{eqnarray}
is the integral over $M$ of a local function evaluated for sections
$\phi$ of $E$. (Of course, we must restrict $M$ and $\phi$ or both for
this to make sense.)
\end{definition}

The variational approach is to seek the critical points of such a local
functional. More precisely, we seek sections $\phi$ such that $\delta
{\mathcal L}[\phi]=0$ where $\delta$ denotes the variational derivative
corresponding to an `infinitesimal' variation: $\phi \mapsto \phi
+\delta\phi$.  For sections of compact support,
the condition $\delta {\cal L}[\phi]=0$ is equivalent to
the Euler-Lagrange equations on the corresponding local function $L$ as
follows: Let
\begin{eqnarray}
D_i={\partial\over\partial x^i}+u^a_{Ii}{\partial\over\partial
u^a_{I}}
\end{eqnarray}
be the total derivative acting on local functions (note that
$D_iu^a_I = u^a_{Ii}$) and
\begin{eqnarray}
E_a=(-D)_I{\partial\over\partial u^a_{I}}
\end{eqnarray}
the Euler-Lagrange derivative.  
(Summation over repeated indices, one up, one down, is understood.)
The notation $(-D)_I$ means
\begin{eqnarray}
\frac{\partial }{\partial u^a}-\partial_i\frac{\partial
}{\partial u^a_i}+\partial_i\partial_j\frac{\partial }{\partial
u^a_{ij}}-....
\ee
The Euler-Lagrange equations are then
\begin{eqnarray}
E_a(L)=0.
\end{eqnarray}

Since $\cal L$ is the integral of an $n$-form on $J$, it is not
surprising that this all makes sense in the deRham complex
$\Omega^*(J)$, which remarkably splits as a bicomplex (though the finite
level complexes $\Omega^*(J^pE)$ do not) \ci{vino,tsujishita,ian:book}.  
The appropriate 1-forms in
the fibre directions are not the $du^a_I$ but rather the {\bf contact
forms} $\theta^a_I= du^a_I- u^a_{Ii}dx^i $. A typical basis element of
$\Omega^{p,q}(J)$ is of the form $f dx^{i_1}\wedge\dots\wedge dx^{i_p}
\wedge \theta^{a_1}_{J_1}\wedge\dots\wedge\theta^{a_q}_{J_q}$ where
$f\in Loc(E)$.  The total differential $d$
splits as $d=d_H+d_V$ where $d_H = dx^iD_i: \Omega^{p,q}\to \Omega^{p+1,q}$ 
and $d_V:\Omega^{p,q}\to \Omega^{p,q+1}$.
We will henceforth
restrict the coefficients of our forms to be local functions, although
we will not decorate $\Omega^*(J)$ to show this.

The Euler-Lagrange derivatives assemble into an operator on forms on $J$:
\begin{eqnarray}
E(L \ dvol_M) = E_a(L)dvol_M\wedge \theta^a 
\end{eqnarray}
where we will usually take $ d{vol}_M = d^nx :=  dx^1\cdots dx^n$.
 
The kernel of the Euler-Lagrange
derivatives is given by total divergences,
\begin{eqnarray}
E_a(L)=0\ \mbox{for all}\ a=1,\dots,k \Longleftrightarrow L=(-1)^k D_i j^i
\end{eqnarray}
for some local functions $j^i$. Equivalently,
\begin{eqnarray}
E_a(L)=0\ \mbox{for all} \ a \Longleftrightarrow L(x,u^{(p)})
d{vol}_M = (-1)^{i-1}d_H  j^i  d{vol}_M/d x^i
\end{eqnarray}
where if $d{vol}_M = dx^1\cdots dx^n,$ then $d{vol}_M/d x^i = dx^1\cdots
dx^{i-1} dx^{i+1} \cdots dx^n$.

A Lagrangian $\cal L$  determines a {\bf stationary surface} or   {\bf
solution surface} or {\bf shell} $\Sigma\subset J^\infty$ such that
$\phi$ is a solution of the variational problem (equivalently, of the
Euler-Lagrange equations) iff $j^\infty \phi$ has its image in $\Sigma$.
The corresponding algebra of local functions, $Loc(\Sigma)$ is isomorphic
to the quotient $Loc(E)/\mathcal I$ where the {\bf stationary ideal} 
$\mathcal I$ consists of
local functions which vanish `on shell', i.e. when restricted to the
solution surface $\Sigma$.  We assume enough regularity that this ideal is 
generated by the local functions $D_IE_a(L)$.
\newcommand\K{\cal K}
\newcommand\KT{\cal KT}
\section{Anti-fields as Koszul generators}
To present my revisionist view of the \bv complex,
let me take you `through the looking glass' and present a `bi-lingual'
(math and physics) dictionary. 

{} From here on, we will talk in terms of algebra extensions of
$\CJ$, but the extensions will all be free graded
commutative. We could instead talk in terms of an extension of $E$ or
$J$ as a super-manifold, the new generators being thought of as
(super)-coordinates.

We extend $\CJ$ by adjoining generators of various
degrees to form a free graded commutative algebra $\BV$ over
$\CJ$, that is, even graded generators give rise to a
polynomial algebra and odd graded generators give rise to a Grassmann
($=$ exterior) algebra. The generators (and their products) are, in
fact, bigraded $(p,q)$; the graded commutativity is  with respect to the
total degree $p-q$. 

We begin by adjoining,
for each variable $u^a_I$, an `anti-field' $u_{aI}^*$ of bidegree $(0,1)$
(again interpret the $u^*_{aI}$ as the formal derivatives of the $u_a^*$)
and thus form the Koszul complex
$$
\mathcal K = Loc(E)\otimes\Lambda u^*_{aI}
$$
where $\Lambda$ denotes the free graded commutative algebra, 
with the derivation differential $\delta$ determined by
$$
\delta u^*_{aI} = D_IE_a(L)
$$
so that $H^0(\mathcal K)\simeq Loc(E)/\mathcal I$.

Notice that we could interpret the pairing of $u^a$ with $u^*_a$
as giving rise to a Poisson-like bracket on $\K$; this will be carried 
out in section 4  and called the `anti-bracket' because of the
signs that intervene.  The Koszul complex will then play a role 
analogous to that of the basic Poisson algebra in the 
Batalin-Fradkin-Vilkovisky approach to the cohomological reduction of 
Hamiltonian systems with first class constraints \cite{jim:hrcpa,lars:diss}.
\section{Noether identities and Tate generators}
The (left hand sides of the) Euler-Lagrange equations generate $\cal I$ 
as a differential ideal, that is, as an ideal over the ring of 
total differential operators,
but this means that we may have {\bf Noether identities} not only of the form
\begin{eqnarray}
r^aE_a(L) = 0
\end{eqnarray}
but also
\begin{eqnarray}
r^{a I}D_IE_a(L) = 0, \label{ni}
\end{eqnarray}
where $r^{aI}\in Loc(E)$.
Of course we have `trivial' identities of the form
\begin{eqnarray}
D_JE_b(L)\mu_\alpha^{bJaI}D_IE_a(L)=0,
\end{eqnarray}
with $\mu_\alpha^{bJaI}$ skew-commutative in the index pairs $bJ$ and $aI$
since we are dealing with a commutative algebra of functions.  

We now assume we have a set $\{\alpha\}$ of indices such that the 
corresponding identities 
\begin{eqnarray}
r_\aaa := r^{aI}_\aaa D_I E_a(L) = 0,
\end{eqnarray}
with $r^{aI}_\alpha \in Loc(E),$ generate all the non-trivial relations 
in $\cal I$, as a differential
ideal, that is, over the algebra of total differential operators
$s^{I} D_I$. According
to Noether \cite {noether}, each such identity corresponds to a 
family of {\bf infinitesimal  gauge symmetries} depending on
arbitrary local functions in $Loc(E)$, i.e. infinitesimal variations
 that preserve the space of solutions up to total divergences,
or, equivalently,  vector fields
tangent to $\Sigma$. (Noether considers only the dependence on 
functions in $Loc(M)$.)
For each Noether identity indexed by $\alpha$, we
denote the corresponding family of vector fields by $\delta_\alpha (f)$. 
We denote by
$\Xi$, the {\bf space  of gauge symmetries}, considered as a vector
space but also  as a module over $\CJ$.

For electricity and magnetism and, more generally, Yang-Mills theory,
the fields are the components of the gauge potential, i.e. of a Lie
algebra valued connection for a $G$-bundle over $M = \ Space-Time$.
In local coordinates, $ A = A^a_\mu T_a dx^\mu$ 
where the $T_a$ form a basis for the Lie
algebra $\mathfrak g$.  The Lagrangian is $L=tr(F\wedge *F)$ where $F$
is the curvature (field strength) of $A$, $*$ is the Hodge 
operator with respect to a Lorentz metric on the 4-dimensional $M$
and $tr$ denotes trace.
The corresponding Euler-lagrange equations are (the components of)
$$D_A*F=0,$$
where $D_A$ is the covariant derivative corresponding to the connection $A$.
The Noether identities are  (the components of) $$D_AD_A*F=0$$.
The corresponding {\it infinitesimal}
gauge symmetries are
$$A\mapsto A + d_Af$$
where $f$ is an arbitrary $\mathfrak g$ valued function on $M$.
In local coordinates,
$$A^a_\mu \mapsto A^a_\mu + \partial_\mu f^a + C^a_{bc} A^b_\mu f^c.
$$
The existence of non-trivial Noether identities implies that $H^1(\mathcal K)$
is not $0$.  To get rid of this unwanted cohomology, Tate \cite{tate} directs
us to adjoin further generators: for each
$r_\alpha$,  adjoin a corresponding 
`anti-ghost' $C^*_\alpha$ of bidegree $(0,2)$ 
and its `derivatives' $C^*_{\alpha I}$
with 
$$\delta C^*_{\alpha I} = D_I r^{aJ}_\alpha u^*_{a J}.$$
In the reducible case, adjoin further variables of bidegree $(0,q)$
so that ultimately the resulting Koszul-Tate complex $\mathcal{KT}$
has 
\be
H^0(\mathcal{KT})= Loc(E)/ \mathcal I\\
H^i(\mathcal{KT}) = 0 {\rm \ for\ } i\neq 0.
\ee

\section{Ghosts and the anti-bracket}
Now for each $r_\alpha$,  further adjoin a corresponding ghost $C^\alpha$ 
and derivatives $C^\alpha_I$ of bidegree $(1,0)$.
For all these extended jet variables, we again have the obvious
actions of the $D_i$, e.g. $D_iC^\alpha_I =  C^\alpha_{Ii}.$

The resulting algebra, due to Batalin and Vilkovisky,  we denote $\BV.$
 Here is a table showing the relevant math terms and the bidegrees.

\medskip

\begin{tabular}{|l|l|c|c|c|}
\hline
Physics & Math & Ghost & Anti-ghost & Total\\[-2pt]
Term & Term & Degree & Degree & Degree \\
\hline
field & section & 0 & 0 & 0 \\[6pt]
anti-field & Koszul generator & 0 & 1 & -1 \\[6pt]
ghost & Chevelley-Eilenberg  generator & 1 & 0 & 1 \\[6pt]
anti-ghost & Tate  generator & 0 & 2 & -2 \\
\hline
\end{tabular}

\medskip

\noindent Note that the anti-field coordinates depend on $E$ but the
ghosts and anti-ghosts depend also on the specific Lagrangian.

Following Zinn-Justin and Batalin-Vilkovisky,
 this algebra in turn can be given an 
{\bf  anti-bracket} $(\ ,\ )$ of
degree $-1$ which, remarkably, combines with the product we began with
to produce a very strong analog of a Gerstenhaber
algebra (see section 9), although this was not recognized until quite
recently. (Gerstenhaber's first example occurs in \cite{gerst:coh}
but the name is more recent \cite{lz:new,kvz}.)
The anti-bracket is so called because of the degree shift and
the corresponding signs for the symmetry under interchange
of the two entries (cf. the Whitehead product in homotopy theory). 
 In the original field-anti-field formalism, the anti-bracket was 
often expressed in terms of Dirac delta-functions.  In the jet bundle 
setting, the anti-bracket can be defined as follows: First, to simplify signs, 
physicists use both left and right graded derivations of graded algebras. By 
this is meant that $\theta$ is a left derivation if
$\theta (ab) = \theta(a)b + (-1)^{|\theta||a|}a\theta (b)$ and $\rho$ is a
right derivation if $\rho(ab) = a\rho(b)+(-1)^{|\rho||b|}\rho (a)b.$ 
In keeping with the tradition in algebraic topology, we will not decorate left 
derivations, but will indicate the corresponding right derivation by
$\bar\theta$ defined by $\bar\theta (a) = (-1)^{|\theta||a|} \theta a.$
Now define the anti-bracket $(\ ,\ ):\BV\otimes\BV\to \BV$ by 
$$
(A,B) = E_a(A)\bar E_*^a(B) - E_*^a(A)\bar E_a(B) 
+ E_\alpha (A)\bar E_*^\alpha (B) - E_*^\alpha (A)\bar E_\alpha (B) .
$$
\noindent
Here the extended Euler-Lagrange derivatives 
$E_*^a,E_\alpha, E_*^\alpha$ are defined by the formal analogs  for 
the variables $u^*_a, C^\alpha, C^*_\alpha$ of the formula for $E_a.$
The only non-zero anti-brackets of generators
are
$$
(u^a,u_b^*) = \delta^a_b \ \ {\mbox {and}\ \ } (C^\alpha, C^*_\beta)
=\delta^\alpha_\beta,$$ and $(A, \ )$ acts as a graded derivation.

In light of this odd symplectic pairing, it is $\K$ rather 
than $\CJ$ that plays the role of the basic
Poisson algebra in the homological reduction of constrained 
Hamiltonian systems, though here it is an odd analog.
\def\cc{ f^\alpha_{\bb\ggg}C^\bb C^{\ggg}}
\section{The Batalin-Vilkovisky complex $(\BV,s_\infty)$}
Batalin and Vilkovisky use the anti-bracket to make their algebra into
a differential graded algebra using the anti-bracket and the 
Lagrangian as follows:

Define an operator $s_0$ of degree $-1$ on $\BV$ as $ (L,\ )$.

This reproduces the Koszul complex in that
\begin{eqnarray}
s_0 u^*_a =  {{\dd L}\over {\dd u^a}}
\end{eqnarray}
as in the Koszul complex for the ideal,
so that $H^{0,0}\subset \CJ$ is given by 
${{\dd L}\over {\dd u^a}} = 0,$
but  
\begin{eqnarray}
s_0(r^a_\alpha u^*_a) = r^a_\alpha  {{\dd L}\over {\dd u^a}} =0,
\end{eqnarray}
so that $H^{0,1} \neq 0.$

Now consider the extended Lagrangian $L_0+L_1$ with
\begin{eqnarray}
L_1 =  u^*_a (-D)_I [r^{aI}_\alpha C^\alpha]
\end{eqnarray}
and $s_1 = (L_0+L_1,\ ),$
so that
\begin{eqnarray}
s_1 C^*_\alpha = u^*_a r^a_\alpha
\end{eqnarray}
reproducing the Koszul-Tate differential.
To begin to capture the Chevalley-Eilenberg differential,
further extend $L_0+L_1$ by adding a term
\begin{eqnarray}
L_2 =  C_\alpha^*f^{IJ\alpha}_{\beta\gamma}D_IC^\beta D_JC^\gamma,
\end{eqnarray}
where the $f$ represent  a bidifferential operator.  In the special case
in which the $f^\alpha_{\beta\gamma}$ and the $r^a_\alpha$ 
are constants, we have,
 for $s_2 = (L_0+L_1+L_2,\ ),$
$$
s_2C^\alpha = f^\alpha_{\bb\ggg}C^\bb C^{\ggg}
$$
$$
s_2u^a = \rr C^\alpha,
$$
which is how the Chevalley-Eilenberg coboundary looks in terms of bases
for a Lie algebra and a module and corresponding structure constants. 
However, in general, we may not have $(s_2)^2 = 0$ 
since $\rr$ and $f^\alpha_{\bb\ggg}$
are, in general,  functions.
Batalin and Vilkovisky prove that all is not lost:

\begin{thm} $L_0+L_1+L_2$ can be further extended by terms of higher degree in the
ghosts to $L_\infty$  so that $(L_\infty,L_\infty) = 0$ and
hence the corresponding $s_\infty$ will have square zero.
\end{thm}

With hindsight, we can see that the existence of these terms of higher
order is guaranteed because the antifields and antighosts provide a
resolution of the stationary ideal. The pattern of the proof is one
standard in Homological Perturbation Theory 
\cite{victor:pert,Gugmay,gugjds,GLS,joh:small,HK}: If $s_n^2\neq 0,$
at least $\delta s_n^2 = 0$, so by the acyclicity of $\mathcal KT$, there
exists $L_{n+1}$ as desired. (See \cite{jim:hrcpa,lars:diss} for details in the
corresponding Hamiltonian case.)

We refer to this complex $(\BV,    s_\infty)$ as the {\bf \bv complex}.
The differential $s=s_\infty$ is sometimes called the BRST operator and
its cohomology the BRST cohomology. Physicists tend to write $H^*(s)$
where mathematicians would write $H^*(\BV).$

What is the \bv complex computing, that is, what do the groups $H^i$
mean for $i> 0$?  Given sufficient regularity conditions on the
Euler-Lagrange equations $E_a(L)=0$, there is a nice
geometric answer \cite{ht, barnich:diss}, again analogous to the 
Hamiltonian case.  The stationary
surface $\Sigma$ is foliated by the gauge orbits of the gauge symmetries.
The standard complex for the longitudinal cohomology, i.e. the
de Rham cohomology using all of $Loc(\Sigma)$ but differential forms only along the leaves (gauge orbits) of the foliation is given by 
$Alt_{Loc_E/{\cal I}}(\Xi, Loc_E/{\cal I})$ with the Chevalley-Eilenberg 
differential. In the regular     case, the \bv complex has replaced 
$Loc_E/{\cal I}$ by the Koszul-Tate resolution and $\Xi$ by a corresponding
resolution.
\section{The BV version of the \varbic }
In the presence of a fixed Lagrangian with or without gauge symmetries, there 
are various ways of combining the \varbic with the BV approach. The simplest is to 
start with the usual \varbic and adjoin anti-fields, ghosts and anti-ghosts, 
etc., just as we did to $\CJ$.  The exterior derivative acts  in the \varbic as 
before and trivially on the new variables.  A more subtle way is to treat the 
new variables $u^*_a, C^\alpha, C^*_\alpha$ as true coordinates on a 
super-manifold $E\times {\mathbf R}^{j|j+k}$ with corresponding jet coordinates.   Then we would have the corresponding \varbic, meaning that we would have
differentials $du^*_a,$ etc. and corresponding contact forms.  

 An intermediate 
alternative is to consider only $u^*_a$ and $C^*_\alpha$ as
super-coordinates and leave the ghosts as just Chevalley-Eilenberg generators 
for computing the equivariant cohomology for the appropriate Lie algebra or 
$L_\infty$-algebra.

This intermediate alternative should be compared with the techniques
of rational (or real) homotopy theory (which have also made an appearance
in the physics literature, for example in 
Vasiliev\cite{vasiliev:consistent,vasiliev:prop}).
There one would construct a `model' of $\Omega^*(\Sigma)$ of the form
$\Omega^*(J)\otimes \Lambda Z$ for a graded vector space $Z$ determined 
essentially by the same process as the Koszul-Tate resolution for $Loc(\Sigma)$.

Question: Does the  complex $\Omega^*(J)\otimes \Lambda(u^*_{aI},
C^*_{\alpha I})$ give a real homtopy theory model for $\Omega(\Sigma)$?

To date, none of these alternatives have been completely 
fleshed out, though certain 
computations in the physics literature may allow for one or another of these 
interpretations \cite{mccloud, bh:isoms}.

What is the significance of $(s_\infty)^2 = 0$ in our Lagrangian
context  or, equivalently, of the {\bf Master Equation} $(\L, \L) = 0$?
 There are three answers:  in higher
homotopy algebra, in deformation theory and in mathematical physics.

\section{The Master Equation and Higher Homotopy Algebra}

As Henneaux has emphasized, all the structure of our problem - the
Noether identities, the gauge symmetries, the reducibilities, etc. - have 
all been encoded in $s_\infty$.
If we expand $s = s_0 + s_1 + ...$, where the subscript indicates the
change in the ghost degree $p$, the individual $s_i$ do not correspond
to $(L_j, \ )$ for any term $L_j$ in $\L$ but do have the following
graphical description in terms of the bigrading $(p,q)$:
\input{epsf}
\begin{center}
\mbox{
\epsfxsize=1.6in
\epsfbox{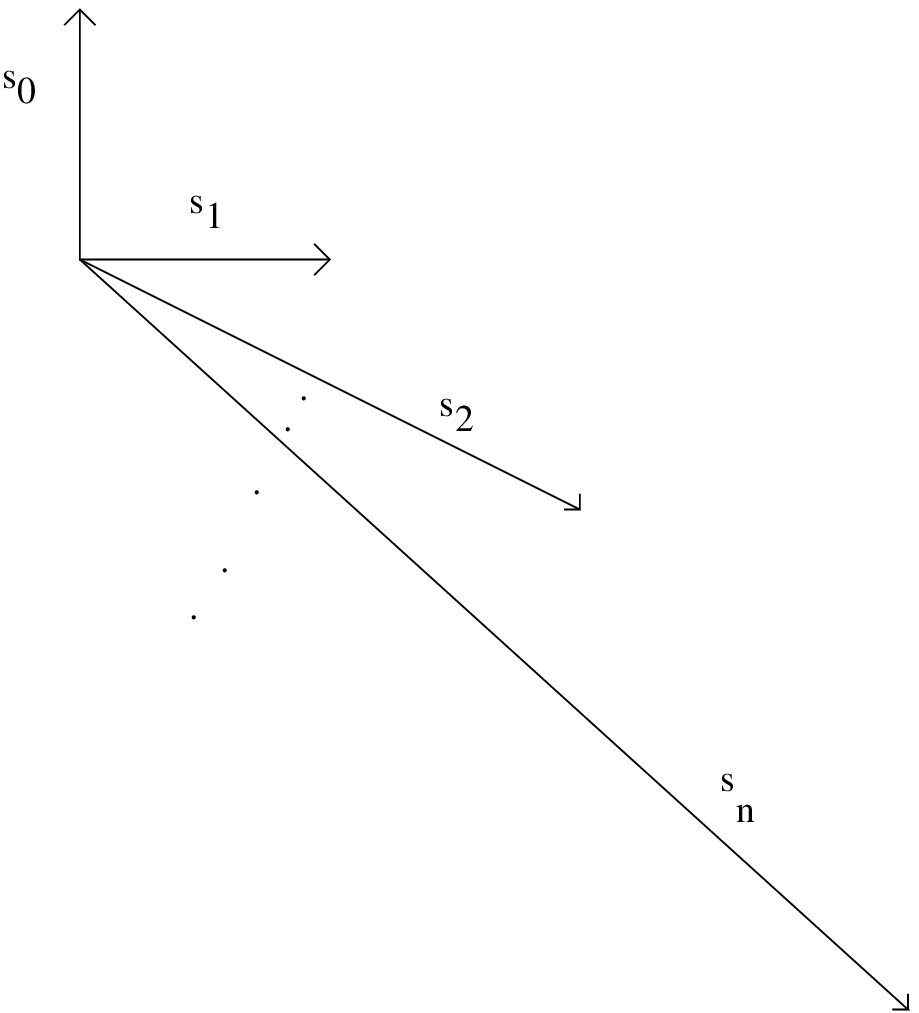}
}
\end{center}
\noindent so that we see the BV-complex as a multi-complex. The
differential $s_0$ gives  us  the Koszul differential $\delta$, while
$s_1$ gives us the Koszul-Tate $\delta$ and
part of $s_2$ looks like that of Chevalley-Eilenberg.  That is,
$C_\alpha^* f^\alpha_{\bb\ggg}C^\bb C^{\ggg}$ describes a
(not-necessarily-Lie) bracket.  Indeed, if the structure functions 
for $L_1$ and $L_2$ are in fact constants, we could have exactly the
Chevalley-Eilenberg cochain complex for a Lie algebra with coefficients in 
a module; in our case, the module is the Koszul-Tate resolution.
But for structure functions, we need the terms of higher order.
What is the significance of, for example,
terms with one anti-field $u^*_a$, one anti-ghost
$C_\alpha^*$ and three ghosts $C^\bb C^{\ggg} C^\delta$? Such terms
 describe a tri-linear $[\ ,\ ,\ ]$  on gauge symemtries
and so on for multi-brackets of possibly
arbitrary length.  Moreover, the graded commutativity of the underlying
algebra of the BV-complex implies appropriate symmetry of these
multi-brackets.  

It is here that I began to see the shadow of a {\bf
strong homotopy Lie algebra (sh Lie algebra)} or  $L_\infty$ {\bf algebra}
\cite{ls}. 

A Lie algebra is a vector space $V$ with a skew map $V\otimes V\to V$ 
satisfying the Jacobi identity; this is equivalent to a coderivation $D$ 
of degree $-1$ on the graded
commutative {\it co}algebra on $V$ regarded as in degree 1 with $D^2=0$.

A graded Lie algebra is a graded vector space $V=\oplus V_i$ with  
graded skew maps $V_i\otimes V_j\to V_{i+j}$ satisfying 
the Jacobi identity with appropriate signs; this can be encoded   as  a 
coderivation $D$ of degree $-1$ on the graded
commutative coalgebra $\Lambda\uparrow V$ on $\uparrow V$ 
with $D^2=0$
(where $\uparrow V$ denotes the graded vector space obtained from$V$
by shifting the grading up by 1).

Certain aspects of the algebra we are concerned with are much easier to handle
after this shift and extension to the graded symmetric algebra.  Vasiliev
has begun the study of higher spin interactions at this level 
\cite{vasiliev:consistent,vasiliev:prop}.
Since the \bv algebra includes 
 a coderivation $s$ on a graded commutative coalgebra,
the condition that $s_\infty^2=0$ translates to include the
following identities, which are the defining identities for an
$L_\infty$-algebra \cite{ls}:
\begin{eqnarray}
d[v_1, \dots, v_n] + \sum_{i=1}^n  \epsilon(i) [v_1, \dots, dv_i, \dots ,
v_n] \nonumber\\
= \sum_{ p+q = n+1}
\sum_{ \mbox{unshuffles } \ \sigma}
\epsilon (\sigma) [[v_{i_1}, \dots, v_{i_p}], v_{j_1}, \dots, v_{j_{q}}],
\end{eqnarray}
where $\epsilon (i) = (-1)^{|v_1| + \dots + |v_{i-1}|}$ is the sign
picked up by taking $d$ through $v_1,  \dots,  v_{i-1}$ and, for the
unshuffle $\sigma:\{1,2, \dots, n\}\mapsto  \{i_1, \dots, i_k, j_1,
\dots, j_{l}\}$ with $i_1 < \dots < i_p$ and $j_1<\dots < j_q$, 
the sign $\epsilon (\sigma)$ is the sign picked up by
the $v_i$ passing through the $v_j$'s during the unshuffle of
$v_1, \dots , v_n$, as usual in superalgebra.

Realized in the \bv complex,
 these defining identities tell us, for small values of $n$, that $d_{KT}
$
is a graded derivation of the bracket, that the bracket may not satisfy the
graded Jacobi identity but that we do have (with the appropriate signs)
\begin{eqnarray}
[[v_1, v_2], v_3] \pm [[v_1, v_3], v_2] \pm [[v_2, v_3], v_2]
=\hspace{1.7in}\nonumber
\\
\hspace{.5in}- d[v_1, v_2, v_3] \pm [dv_1, v_2, v_3]\pm[v_1, dv_2,
v_3]\pm [v_1, v_2, dv_3].
\end{eqnarray}
i.e. the Jacobi identity holds {\it up to homotopy} or, for closed
forms, the Jacobi identity holds modulo an exact term given by
 the tri-linear bracket.

Note that the identity is the Jacobi identity if $d_{KT}=0$.
Although the other brackets are then not needed for the Jacobi identity
(we have a strict graded Lie algebra), the other brackets need
not vanish and indeed such brackets occur in the homology of a dg Lie 
algebra and are known as Massey-Lie brackets \cite{retakh}.
 (They are the Lie analogs
of the Massey products first introduced in cohomology of topological spaces
\ci{massey,massey-uehara}.)

Furthermore, the identity has content even if only for one $n$
is the $n$-linear bracket non-zero, all others vanishing.  Precisely
that situation has recently been studied quite independently of my work
and of each other by Hanlon and Wachs \cite{hanlon-wachs} (combinatorial
algebraists), by Gnedbaye \cite{gned} (of Loday's school) and by
Azcarraga and Bueno \cite{az-bu} (physicists). 
On the other hand, this is not Takhtajan's fundamental identity
\cite{takh:nambu} for the generalized Nambu $n$-linear `bracket'.
(This identity was known also
to Flato and Fronsdal in 1992, though unpublished, and to Sahoo and Valsakumar
\cite{sahoo}.)

See M. M. Vinogradov's talk in these proceedings 
for a comparison of these two distinct generalizations of
the Jacobi identity for $n$-ary brackets (including reference to
Filippov \cite{filippov}).

\section{Deformation Theory and the Master Equation in Field Theory}

In the Lagrangian setting, we wish to deform not just the  local
functional, but rather the underlying local function $L$. In the case of
electricity and magnetism (Maxwell's equations) and Yang-Mills, 
 the relevant algebra of gauge symmetries is described by a
finite dimensional Lie algebra which, moreover, holds off shell. In
terms of an appropriate basis and in the notation of section 3 for
families of vector fields, we have
\begin{eqnarray}
\lbrack\delta_\alpha,\delta_\beta\rbrack=
f^\ggg_{\aaa\bb}\delta_{\gamma}
\end{eqnarray}
for structure {\it constants} $f^\gamma_{\alpha\beta}$ and acting
on all fields, not just on solutions. This allows the extended Lagrangian
to be no more than quadratic in the ghosts. 

As field theories, a Yang-Mills `particle' can be described by a field of 
spin 1, while  the graviton   can be described by
a field of spin 2.  Somehow (Mother Nature?)
this is related to the strict Lie algebra
structures just described. For higher spin particles, however, we have
quite a different story, which first caught my attention in the work of
Burgers, Behrends and van Dam \cite{burgers:diss,BBvD:three}, though I have 
since
learned there was quite a history before and after that and major questions 
still remain open.  By higher spin particle Lagrangians, I mean that the
fields are symmetric $s$-tensors (sections of the symmetric $s$-fold
tensor product of the tangent bundle). If the power is $s$, the field is
said to be of spin $s$ and represents a particle of spin $s$. \BBvD \
start with a free theory with abelian gauge symmetries and calculate all
possible infinitesimal (cubic) interaction terms up to the appropriate
equivalence (effectively calculating the appropriate homology group,
as in formal deformation theory).
They then sketch the problem of finding higher order terms for the
Lagrangian, but do not carry out the full calculation.  In fact,
according to the folklore in the subject, a consistent theory for $s\geq
3$ will require additional fields of arbitrarily high spin $s$.  For
$s=3$, the statement is that all higher integral spins are needed,
while for $s=4$, all higher even spins.  

From the deformation theory point of view, we are deforming the
pair consisting of the abelian Lie algebra and the stationary surface
while constraining the stationary surface to remain of Lagrangian type.
Stephen Anco \cite{anco:constr,anco:spin1,anco:threehalves}
 has looked at deformations of the gauge symmetries and the Euler-Lagrange
equations directly as opposed to our interpretation of \bv as deforming
the Lagrangian function.  
In contrast, Krasil'shchik \cite{krasil} considers deforming 
the stationary surface
(more generally, the diffiety) without regard to gauge symmetries.
Deformation theory \cite{gerst:defm} suggests the following
attack:  Compute the primary obstructions and discover that all
infinitesimals are obstructed.  Add additional fields to kill the
obstructions and calculate that indeed additional fields of arbitrarily
high spin $s$ are needed. Whether such computations are feasible 
to all orders remains to be seen; perhaps eventually they lie
in cohomology groups which can be seen to vanish. In one memorable phrase, 
this would be `doing string field theory the hard way''.

In contrast, Zwiebach \cite{z:csft} does indeed have a consistent closed 
string field
theory (CSFT), but produced in an entirely different way.  Recall one of
the earliest examples of deformation quantization, the Moyal bracket.
Moyal was able to produce a non-trivial deformation of the  commutative
algebra of smooth functions on $\mathbb R^{2n}$, with infinitesimal given by
the  standard Poisson bracket, by writing down the entire formal power 
series.\footnote{Since the conference, Kontsevich \cite {kont:defquant} 
has constructed a
deformation quantization of the algebra of smooth functions on any
Poisson manifold with infinitesimal given by the Poisson bracket.
In a remarkable tour de force, he produces the
entire formal power series via a specific $L_\infty$-map to a sub dg Lie algebra
of the Hochschild cochain complex from its homology.}
Similarly, Zwiebach is able to describe the entire CSFT Lagrangian (at
tree level) by giving it in terms of the differential geometry of the
moduli space of punctured Riemann spheres (tree level $=$ genus $0$).
In fact, Zwiebach has the following structure: a differential graded
Hilbert space $(\cal{ H}, <\ ,\ >, Q)$ related to the geometry of the
moduli spaces from which he deduces $n$-ary operations $[\ ,\dots,\ ]$
which give an $L_\infty$ structure.
The deformed Lagrangian (still classical) and hence the Master Equation
is satisfied  for
$$
S(\Psi ) = {1\over 2} \langle \Psi , Q\Psi \rangle + \sum_{n=3}^\infty
{\kappa^{n-2} \over n!} \{ \Psi\dots\Psi \}.
$$
The expression $\{ \Psi\dots\Psi \}$ contains n-terms and will be
abbreviated $\{ \Psi^n \};$ it is given in terms of the brackets by $\{
\Psi\dots\Psi \} = \langle \Psi , [ \Psi,\dots, \Psi ] \rangle.$

The field equations follow from the classical action by simple variation:
$$
\delta S = \sum_{n=2}^\infty {\kappa^{n-2} \over n!} \{ \delta\Psi ,
\Psi^{n-1} \}
$$
with gauge symmetries given by
$$
\delta_\Lambda \Psi = \sum_{n=0}^\infty { \kappa^n \over n!} [\Psi^n,
\Lambda ] .
$$

Since the conference, Kontsevich \cite {kont:defquant} has constructed a
deformation quantization of the algebra of smooth functions on any
Poisson manifold with infinitesimal given by the Poisson bracket.
In a remarkable tour de force, he produces the 
entire formal power series via a specific $L_\infty$-map to a sub dg Lie algebra
of the Hochschild complex from its homology.
\section{Quantization and other puzzles}

So far our description of the anti-field, anti-bracket formalism has been
in the context of deformations of `classical' Lagrangians.  Batalin and
Vilkovisky (as well as much of the work on BRST cohomology) were
motivated by problems in quantization. (The `finite dimensionality'
of the BV complex removes some of the problems with the path integral 
approach to quantization.)  The quantum version of the
anti-field, anti-bracket formalism involves a further `second order'
differential operator $\DD$ of square $0$ on the BV complex $\BV$
relating the
graded commutative product and the bracket - namely, the bracket is the
deviation of the operator $\DD$ from being a derivation of the product. 
This has led to the abstract definition of a BV-algebra as a Gerstenhaber
algebra with additional structure \cite {lz:new}.

\begin{df}\rm  A {\bf Gerstenhaber algebra} is a graded commutative algebra
$A$ with a bracket $[\ , \ ]$ of degree $-1$ satisfying the graded Jacobi
identity, i.e $[a,\ ]$ is a graded derivation of the bracket,
and $[a,\ ]$ is also a graded derivation of the commutative product:
$$
[a,bc] = [a,b]c + (-1)^{|b|(|a|+1)} b[a,c].
$$
\end{df}

\begin{df}\rm  A {\bf BV-algebra} is a Gerstenhaber algebra with an
operator (necessarily
of degree $-1$ for a bracket of degree $-1$) such that
$$
[A,B] = \DD(AB) -\DD(A)B +(-1)^A\DD(B).
$$
\end{df}

Alternatively, a definition can be given in terms of a graded commutative
algebra with an appropriate  `second order'
operator $\DD$ with the bracket being defined by the above equation \cite{
schw:bvgeom,akman:diffop}.

The quantization of Zwiebach's CSFT involves further expansion of the
Lagrangian in terms of (the moduli space of) Riemann surfaces of genus
$g\geq 0.$  Here the operator $\DD$ is determined by the self-sewing of a
pair of pants (a Riemann sphere with 3 punctures).  Now Zwiebach's CSFT
provides a solution of the `quantum Master Equation' which
in the context of a BV-algebra, is
$$
(S,S)=1/2 i\hbar\DD S.
$$
Again we see an analog of the Maurer-Cartan equation or of a flat
connection, but why?

\newcommand{\etalchar}[1]{$^{#1}$}
\providecommand{\bysame}{\leavevmode\hbox to3em{\hrulefill}\thinspace}

\end{document}